\documentclass[fleqn,twoside]{article}
\usepackage[headings]{espcrc2}

\readRCS
$Id: espcrc2.tex,v 1.2 2004/02/24 11:22:11 spepping Exp $
\usepackage{graphicx}
\usepackage[figuresright]{rotating}

\newcommand{\ra}{\rightarrow}
\newcommand{\llb}{l^+ l^-}

\newcommand{\AmS}{{\protect\the\textfont2
  A\kern-.1667em\lower.5ex\hbox{M}\kern-.125emS}}

\title{New physics upper bound on the branching ratio of
$B_s \ra l^+ l^-$ and $B_s \ra l^+ l^- \gamma$.}

\author{Ashutosh Kumar Alok \address[IITB]{Department of Physics, 
Indian Institute of Technology, Bombay, \\
        Mumbai 400076, India} and
S. Uma Sankar\addressmark[IITB]}

\runtitle{New physics upper bound on the branching ratio of 
$B_s \ra l^+ l^-$ and $B_s \ra l^+ l^- \gamma$.}
\runauthor{A. K. Alok and S. Uma Sankar}

\begin{document}

\begin{abstract}
We consider the most general new physics effective Lagrangian for
$b \rightarrow s l^+ l^-$. We derive the upper limit on the branching
ratio for the processes $B_{s}\rightarrow l^{+}l^{-}$ where $l=e,\,\mu$,
subject to the current experimental bounds on related processes,
$B \rightarrow (K,K^*) l^+ l^-$. If the new physics interactions are of
vector/axial-vector form, the present measured rates for $B \rightarrow
(K,K^*) l^+ l^-$ constrain $B(B_{s}\rightarrow l^{+} l^{-})$ to be of
the same order of magnitude as their respective Standard Model (SM) 
predictions. On the other hand, if the new physics interactions are of
scalar/pseudoscalar form, $B \rightarrow (K,K^*) l^+ l^-$ rates do not
impose any useful constraint on $B(B_{s}\rightarrow l^{+} l^{-})$ and
the branching ratios of these decays can be as large as present
experimental upper bounds. If future experiments measure
$B(B_{s}\rightarrow l^{+} l^{-})$ to be $\geq 10^{-8}$ then the new physics
giving rise to these decays has to be of the scalar/pseudoscalar form.
We also consider the effect of new physics on 
$B(B_s \rightarrow l^+ l^- \gamma)$ subject to the
present experimental constraints on $B \rightarrow (K,K^*) l^+ l^-$
and $B \rightarrow K^* \gamma$. New physics in form scalar/pseudoscalar,
which makes a very large contribution to $B_s \rightarrow l^+ l^-$,
makes {\it no contribution at all} to $B_{s}\rightarrow l^{+}l^{-}\gamma$ 
due to angular momentum conservation. New Physics in the
form of vector/axial-vector operators is constrained by the data
on $B \rightarrow (K,K^*) l^+ l^-$ and new physics in the form of
tensor/pseudo-tensor is constrained by the data on
$B \rightarrow K^* \gamma$. In both cases, enhancement of
$B(B_{s}\rightarrow l^{+}l^{-}\gamma)$ much beyond the SM
expectation is impossible. In conclusion, present data on 
$B \rightarrow(K,K^*)$ transitions allow for 
large $B(B_s \rightarrow l^+ l^-)$ but do not allow
$B(B_s \rightarrow l^+ l^- \gamma)$ to be much larger than its
SM expectation.

\vspace{1pc}
\end{abstract}

\maketitle

\section{INTRODUCTION}

The rare decays of $B$ mesons involving flavour changing neutral interactions
(FCNI) $b\ra s$ have been a topic of great interest for long. Not only will 
they subject the Standard Model (SM) to accurate tests but will also 
put strong constaraints on several models beyond the SM.
Recently, the very high statistics experiments at
B-factories have measured non-zero values for the branching
ratios for the FCNI processes $B\rightarrow(K,K^*)l^+l^-$
\cite{babar-03,belle-03},
{\setlength \arraycolsep{1pt}
\begin{eqnarray}
B(B\rightarrow Kl^{+}l^{-}) &=&
(4.8_{-0.9}^{+1.0}\pm0.3\pm0.1)\times10^{-7}, \nonumber \\
B(B\rightarrow K^{*}l^{+}l^{-}) &=&
(11.5_{-2.4}^{+2.6}\pm0.8\pm0.2)\times10^{-7}.\nonumber\\
\end{eqnarray}
These branching ratios are close to the values predicted by the SM
\cite{ali-02}. However, the SM predictions for them contain
about $\sim15\%$ uncertainty coming from the hadronic form factors.
Still, it is worth considering what constraints these measurements
impose on other related processes.

In section 2 and 3 we will discuss the impact of there measurements on 
the predictions for $B_{NP}(B_s \ra l^+ l^-)$ and 
$B_{NP}(B_s \ra l^+ l^- \gamma)$ respectively \cite{aloksankar1,aloksankar2}.

\section{NEW PHYSICS UPPER BOUND ON $B(B_s \ra l^+ l^-)$.}

The same $b \ra sl^+l^-$ four Fermi interaction is responsible for both 
leptonic decays $B_s \ra l^+ l^-$ and semi-leptonic 
decays $B\rightarrow(K,K^*)l^+l^-$. The SM predictions for the branching 
ratios for the decays
$B_s\rightarrow e^+e^-$ and $B_s\rightarrow\mu^+\mu^-$
are $(7.58\pm3.5)\times10^{-14}$ and $(3.2\pm1.5)\times10^{-9}$
respectively \cite{buras-01}. The large uncertainy in the SM prediction
for these branching ratios arises due to the $12 \%$ uncertainty in
the $B_s$ decay constant and $10 \%$ uncertainty in the measurement of
$V_{ts}$. 

$B_s \ra l^+ l^-$ has been studied in various models, both with and 
without natural flavour conservation, before. In both these kinds of 
models it was shown that $B_s\rightarrow\mu^+\mu^-$ 
can have a branching raio of $\geq 10^{-8}$ \cite{london-97,hewett-89}.
From the experimental side, at present, there exist
only the upper bound
$B(B_{s}\rightarrow\mu^{+}\mu^{-}) < 1.0\times10^{-7}$ at $95 \%$ C.L.
\cite{tonelli}.

The effective new physics Lagrangian for $b\rightarrow sl^+l^-$
transitions can be written as,
\begin{equation}
L_{eff}\,(b\rightarrow sl^+l^-)=L_{VA} + L_{SP} + L_T.
\end{equation}
where, $L_{VA}$ contains vector and axial-vector couplings,
$L_{SP}$ contains scalar and psuedo-scalar couplings and
$L_T$ contains tensor couplings. $L_T$ does not contribute
to $B_s \rightarrow l^+ l^-$ because $\langle 0| \bar{s}
\sigma^{\mu \nu} b | B_s(p_B) \rangle = 0$. Hence we
will drop it from further consideration. 
We consider $L_{VA}$ and $L_{SP}$ one at a time.

We parametrize $L_{VA}$ as,
{\setlength \arraycolsep{1pt}
\begin{eqnarray}
L_{VA}\,(b\rightarrow sl^+l^-)&=&\frac{G_{F}}{\sqrt{2}}
\left(\frac{\alpha}{4\pi s_W^2}\right)
\bar{s}(g_V+g_A\gamma_5)\gamma_\mu b \nonumber\\
&~&\bar{l}(g^{'}_V+g^{'}_A\gamma_5)\gamma^\mu l.
\end{eqnarray}
Here the constants $g$ and $g'$ are the effective couplings which
characterise the new physics. The calculation of decay rate gives,
\begin{eqnarray}
\Gamma_{NP}(B_{s}\rightarrow l^{+}l^{-})&=&
\frac{G_{F}^{2}f_{B_{s}}^{2}}{8\pi}
\left(\frac{\alpha}{4\pi s_{W}^{2}}\right)^{2} \nonumber\\
&~&(g_{A}g_{A}^{'})^{2}m_{B_{s}}m_{l}^{2}.
\end{eqnarray}
Thus the decay rate depends upon the value of $(g_{A}g_{A}^{'})^{2}$.
To estimate the value of $(g_{A}g_{A}^{'})^{2}$, we look 
at semi-leptonic decays. We first consider $B\rightarrow K^{*}l^{+}l^{-}$.
The decay rate is,
\begin{eqnarray}
\Gamma_{NP}(B\rightarrow K^{*}l^{+}l^{-})&=&
\frac{1}{2} \left(\frac{G_{F}^{2}m_{B}^{5}}{192\pi^{3}} \right)
\left(\frac{\alpha}{4\pi s_{W}^{2}} \right)^{2} \nonumber\\
&~&(g_{V}^{'2}+g_{A}^{'2})I_{VA}, \label{npk*}
\end{eqnarray}
where $I_{VA}=g_{V}^{2}V^{2}I_{1}+g_{A}^{2}A_{1}^{2}I_{2}$. $I_{1}$ and
$I_{2}$ are integrals over the dilepton invariant mass
($z=q^{2}/m_{B}^{2}$).

$\Gamma_{NP}(B\rightarrow K^{*}l^{+}l^{-})$ depends on both vector 
and axial vector couplings. To get a handle on vector couplings 
we look at $B\rightarrow Kl^{+}l^{-}$.
The decay rate is given by,
\begin{eqnarray}
\Gamma_{NP}(B\rightarrow Kl^{+}l^{-})&=&
\left(\frac{G_{F}^{2}m_{B}^{5}}{192\pi^{3}}\right)
\left(\frac{\alpha}{4\pi s_{W}^{2}}\right)^{2} \nonumber\\
&~&g_{V}^{2}(g_{V}^{'2}+g_{A}^{'2})
\left(\frac{f^{+}(0)}{2}\right)^{2}. \label{npk}
\end{eqnarray}
We are trying to see what is the maximum value of $(g_{A}g_{A}^{'})^{2}$,
consistent with semi-leptonic data. To get this, we make the
approximation $\Gamma_{exp}=\Gamma_{NP}$, i.e. the 
experimentally measuted semi-leptonic branching ratios are 
saturated by new physics couplings. Under this approximation, we get
\begin{equation}
g_{A}^{2}(g_{V}^{'2}+g_{A}^{'2}) = (6.76_{-3.48}^{+4.04})\times10^{-3}.
\end{equation}
Here all the errors were added in quadrature and the values of
form-factors were taken from \cite{ali-00}.

Therefore the upper bounds on the branching ratios are 
$B(B_{s}\rightarrow e^{+}e^{-})<1.20\times10^{-13}$ and
$B(B_{s}\rightarrow\mu^{+}\mu^{-})<5.13\times10^{-9}$ at $3 \sigma$.
These bounds are similar to SM predictions. It should not be surprising 
because $\Gamma = (c.c.)^2 (f.f.)^2 {\rm phase~space}$. In semi-leptonic case
$\Gamma_{exp}=\Gamma_{SM}$. Then we assumed $\Gamma_{NP}=\Gamma_{exp}$ which
implies $(c.c)_{NP}=(c.c)_{SM}$ and hence 
$\Gamma_{NP}(B_s \ra l^+l^-)=\Gamma_{SM}(B_s \ra l^+l^-)$.
A more stringent upper bound is obtained if we equate the new physics 
branching ratio to the difference between the expeimental value 
and the SM prediction. {\it Therefore, given the measured values of
branching ratios of $B\rightarrow(K,K^*)l^+l^-$ by Belle and BaBar, 
new physics cannot boost $B_s \ra l^+l^-$ above SM value if it is of
the form vector/axial-vector}.

Let us turn now to $L_{SP}$ with scalar and pseudoscalar couplings.
\begin{eqnarray}
L_{SP}\,(b\rightarrow sl^{+}l^{-})&=&\frac{G_{F}}{\sqrt{2}}
\left(\frac{\alpha}{4\pi s_{W}^{2}}\right)
\bar{s}(g_{S}+g_{P}\gamma_{5})b \, \nonumber\\
&~&\bar{l}(g_{S}^{'}+g_{P}^{'}\gamma_{5})l.
\end{eqnarray}
The Branching ratio is given by,
\begin{equation}
B(B_{s}\rightarrow l^{+}l^{-})=
0.17\frac{f_{B_{s}}^{2}g_{P}^{2}(g_{S}^{'2}+
g_{P}^{'2})}{(m_{b}+m_{s})^{2}}\label{brbs}.
\end{equation}
To get a bound on $g_{P}^{2}(g_{S}^{'2}+g_{P}^{'2})$ we need to
consider only $B\rightarrow K^{*}l^{+}l^{-}$. Here again we make the 
approximation $\Gamma_{exp}=\Gamma_{NP}$. Under this approximation we get,
{\small
\begin{equation}
g_{P}^{2}(g_{S}^{'2}+g_{P}^{'2})=
\frac{\left(m_{b}-m_{s}\right)^{2}
B_{Exp}(B\rightarrow K^{*}l^{+}l^{-})}{2.16\left[A_{0}(0)\right]^{2}
\times10^{-3}}
\end{equation}}
Substituting this in $(B_s \ra l^+l^-)$ rate we get,
\begin{equation}
B(B_s \ra \mu ^{+} \mu ^{-})=(2 \pm 1) \times 10^{-5}.
\end{equation}
The upper bound on $B(B_s \rightarrow \mu^+ \mu^-)$ from the above
equation is much higher than the present experimental upper bound
\cite{tonelli}. Thus we see that if new phsyics effective Lagrangian is 
of the scalar/pseudoscalar form, then the present measurements of
semi-leptonic rates DO NOT provide any useful constraints on $B_s \ra l^+ l^-$.
{\it Therefore if experiments at Tevatron or LHCb find that 
$B(B_s \rightarrow \mu^+ \mu^-) \geq 10^{-8}$, then we can immediately
conclude that the new phsyics responsible for it is of 
scalar/pseudoscalar type}.

\section{NEW PHYSICS UPPER BOUND ON $B(B_s \ra l^+ l^- \gamma)$.}

We repeated the exercise for $B_s \ra l^+ l^- \gamma$ \cite{aloksankar2}.
The radiative decay
$B_{s}\rightarrow l^{+}l^{-}\gamma$ is free from helicity suppression due
to emission of a photon in addition to the lepton pair. Thus the branching
ratio for this leptonic radiative mode is much higher than that for the
purely leptonic mode despite an additional factor of $\alpha$. 
We are interested on how the current
data on $b \ra s$ transitions, due to the effective interactions
$b \ra s \llb$ and $b \ra s \gamma$, constrain the new physics
contribution to the leptonic radiative decays $B_s \ra \llb \gamma$.

Unlike in the case of $B_{s}\rightarrow l^{+}l^{-}$, if new physics
is in the form scalar/pseudoscalar, then it makes no contribution to
$B_s \ra l^+ l^- \gamma$. The photon has $J=1$. Hence the $\llb$ pair
also must be in $J=1$ state so that the angular momentum of the final
state can be zero. However, by Wigner-Eckert theorem, the matrix element
$\langle \llb (J=1) | \bar{l} (g_s + g_p \gamma_5) l | 0 \rangle$
is zero.

A legitimate question to ask at this stage is: Is it possible to have an order
of magnitude or more enhancement of $B_s \ra l^+ l^- \gamma$ for any type 
of new physics operators? 

We found that if new physics is in the
form of vector/axial-vector operators then the present data on 
$B\rightarrow(K,K^*)l^+l^-$ doesn't allow a large boost for
$B(B_s \ra l^+ l^- \gamma)$. If new phsyics is in the form of 
tensor/pseudotensor operators, then the data on $B\rightarrow(K,K^*)l^+l^-$
gives no useful constraint but the data on $B \ra K^{*} \gamma$ does.
Here again, a large enhancement of $B(B_s \ra l^+ l^- \gamma)$, much 
beyond the SM expectation, is not possible.

{\it Hence we conclude that the present data on $b \ra s$ transitions allow a
large boost in $B(B_s \ra l^+ l^-)$ but not in $B(B_s \ra l^+ l^- \gamma)$, 
compared to SM expectation}.

\section{CONCLUSIONS}
The quark level interaction $b \ra s l^+ l^-$ is responsible for the
three types of decays (a) semi-leptonic $B\rightarrow(K,K^*)l^+l^-$, 
(b) purely leptonic $B_s \ra l^+ l^-$ and also (c) leptonic radiative
$B_s \ra l^+ l^- \gamma$. If $B(B_s \ra l^+ l^-) \geq 10^{-8}$ then the 
new physics operators responsible for this have to be of the form 
scalar/pseudoscalar. Such operators have no effect on 
$B_s \ra l^+ l^- \gamma$. Current data on $B\rightarrow(K,K^*)l^+l^-$
and $B \ra K^{*} \gamma$ do not allow any kind of new physics 
to give rise to a large enhancement of $B(B_s \ra l^+ l^- \gamma)$.

\end{document}